\pdfoutput=1
\documentclass[preprint,preprintnumbers,amsmath,amssymb,floatfix,endfloats*]{revtex4}

\usepackage{graphicx}
\usepackage{dcolumn}
\usepackage{bm}
\usepackage{amsfonts}

\DeclareMathAlphabet{\mathsfsl}{OT1}{cmr}{bx}{it}
\begin{document}
\title{Spatiotemporal analysis of nonaffine displacements in disordered solids sheared across the yielding point}
\author{Nikolai V. Priezjev}
\affiliation{Department of Mechanical and Materials Engineering,
Wright State University, Dayton, OH 45435}
\date{\today}
\begin{abstract}

The time evolution and spatial correlations of nonaffine
displacements in deformed amorphous solids are investigated using
molecular dynamics simulations. The three-dimensional model glass is
represented via the binary mixture, which is slowly annealed well
below the glass transition temperature and then sheared at a
constant strain rate.  It is shown that with increasing strain, the
typical size of clusters of atoms with large nonaffine displacements
increases, and these clusters remain spatially homogeneously
distributed, until the yielding point when mobile atoms become
localized within a system-spanning shear band. Furthermore, the
yielding transition is associated with an abrupt change in the
spatial correlation of nonaffine displacements, which varies from
exponential to power-law decay. We also find that the height of the
first peak in the pair correlation function of small atoms exhibits
a distinct increase at the yielding strain. These results are
discussed in relation to the yielding transition in amorphous
materials under cyclic loading.

\vskip 0.5in


\end{abstract}

\maketitle

\section{Introduction}

The atomic-level understanding of the structure-property
relationship of amorphous materials is important for numerous
structural and biomedical
applications~\cite{BarratRev18,Yucheng19,Qian18}. By now it is well
established that an elementary plastic event in disordered solids
involves a collective rearrangement of a small group of neighboring
particles, or the so-called shear
transformation~\cite{Spaepen77,Argon79}. It was recently argued that
mechanical yield in amorphous solids is a first-order phase
transition from a limited set to a vast number of atomic
configurations~\cite{Procaccia16}.  In the past, the yielding
transition and the strain localization during startup deformation at
a constant rate were repeatedly observed in atomistic simulation
studies~\cite{Varnik04,ShiFalk06,Jacobsen06,Ogata06,Horbach16,
HorbachJR16,Pastewka19}. In particular, it was found that at the
critical strain, the disordered systems exhibit a power-law
distribution of mobile regions, which belongs to the universality
class of directed percolation~\cite{Horbach16}. Remarkably, the
percolation of clusters of nonaffine deformation at the yielding
transition were directly observed in experiments on strained
colloidal glasses~\cite{Schall07}.  More recently, molecular
dynamics simulations of metallic glasses showed the exponential
correlation of nonaffine displacements with a decay length that was
related to the size of shear transformation zones in the elastic
regime of deformation~\cite{Pastewka19}. Although the yielding
phenomenon in deformed disordered systems is ubiquitous, the
complete picture, including the exact form of spatial correlation of
nonaffine displacements and the failure mechanism, remains not fully
understood.

\vskip 0.05in

In recent years, molecular dynamics simulations were widely used to
examine structural relaxation and mechanical properties of amorphous
solids subjected to time periodic deformation~\cite{Priezjev13,
Sastry13,Priezjev14,IdoNature15,Priezjev16,Kawasaki16,Priezjev16a,
Sastry17,Priezjev17,OHern17,Priezjev18,Priezjev18a, NVP18strload,
Regev19,PriMakrho05,PriMakrho09,Kawasaki19,PriezSHALT20}. Most
notably, it was found that in the athermal quasistatic limit,
following a number of subyield cycles, the disordered systems reach
the so-called limit cycle, a dynamic steady state of reversible
particle dynamics, and, in addition, the size of clusters of
particles undergoing cooperative rearrangements becomes comparable
with the system size at the critical strain
amplitude~\cite{IdoNature15}.  During periodic deformation at a
finite temperature, the majority of particles undergo reversible
nonaffine displacements with amplitudes that approximately follow a
power-law distribution~\cite{Priezjev16,Priezjev16a}. Interestingly,
some particles were found to escape their cages temporarily while
still undergoing periodic nonaffine displacements during several
cycles at strain amplitudes below the critical
value~\cite{Priezjev16}.  At sufficiently large strain amplitudes,
the yielding transition occurs after a number of transient cycles
for both slowly and rapidly annealed glasses, and the transition is
associated with the formation of a shear band across the system and
the appearance of a finite hysteresis
loop~\cite{Priezjev17,Priezjev18a}. Despite extensive studies,
however, the precise determination of the critical strain amplitude
and its dependence on the processing history and loading conditions
remains a challenging problem.

\vskip 0.05in

In this paper, the microscopic analysis of particle displacements in
a model glass deformed at constant strain rate is performed using
molecular dynamics simulations. The binary glass is first slowly
cooled well below the glass transition temperature and then strained
above the yielding point. It will be shown that the stress overshoot
occurs at the value of shear strain greater than the critical strain
amplitude for the yielding transition during oscillatory deformation
at the same density and temperature. During shear startup, atoms
with large nonaffine displacements form clusters that are
homogeneously distributed in space, until they suddenly become
localized at the yielding strain. Moreover, the yielding transition
is signaled by a distinct change in the shape of the probability
distribution of the nonaffine measure, spatial correlations of
nonaffinity, and the peak height of the pair correlation function.

\vskip 0.05in

The rest of the paper is structured as follows. The details of
molecular dynamics simulations are given in the next section. The
spatial and temporal analysis of nonaffine displacements, the stress
response, and the pair correlation function are presented in
section\,\ref{sec:Results}.  The results are summarized in the last
section.

\section{Molecular dynamics simulations}
\label{sec:MD_Model}

The molecular dynamics simulations are performed on a well-studied
system of two types of atoms (80:20) that have strongly non-additive
interaction, which prevents crystallization upon
cooling~\cite{KobAnd95}. The binary mixture model was first
introduced and extensively studied by Kob and Andersen (KA) about
twenty years ago~\cite{KobAnd95}. The interaction parameters of the
KA mixture are similar to the parametrization used by Weber and
Stillinger to study the amorphous metal-metalloid alloy
$\text{Ni}_{80}\text{P}_{20}$~\cite{Weber85}. In the KA model, the
interaction between atoms of types $\alpha,\beta=A,B$ is specified
via the Lennard-Jones (LJ) potential, as follows:
\begin{equation}
V_{\alpha\beta}(r)=4\,\varepsilon_{\alpha\beta}\,\Big[\Big(\frac{\sigma_{\alpha\beta}}{r}\Big)^{12}\!-
\Big(\frac{\sigma_{\alpha\beta}}{r}\Big)^{6}\,\Big],
\label{Eq:LJ_KA}
\end{equation}
where the parameters are set to $\varepsilon_{AA}=1.0$,
$\varepsilon_{AB}=1.5$, $\varepsilon_{BB}=0.5$, $\sigma_{AA}=1.0$,
$\sigma_{AB}=0.8$, $\sigma_{BB}=0.88$, and
$m_{A}=m_{B}$~\cite{KobAnd95}. The LJ potential is truncated at the
cutoff radius $r_{c,\,\alpha\beta}=2.5\,\sigma_{\alpha\beta}$ to
alleviate the computational efforts.  In what follows, the
simulation results are reported in the reduced LJ units of length,
mass, energy, and time: $\sigma=\sigma_{AA}$, $m=m_{A}$,
$\varepsilon=\varepsilon_{AA}$, and
$\tau=\sigma\sqrt{m/\varepsilon}$, respectively.  The equations of
motion for each atom were integrated using the velocity-Verlet
algorithm with the time step $\triangle
t_{MD}=0.005\,\tau$~\cite{Allen87,Lammps}.

\vskip 0.05in


We next describe the equilibration procedure and the shear strain
deformation protocol. All simulations were carried out at constant
volume with the corresponding density
$\rho=\rho_A+\rho_B=1.2\,\sigma^{-3}$. The system size is
$L=36.84\,\sigma$ and the total number of atoms is $60\,000$. The
binary mixture was first equilibrated in the liquid state at the
temperature $T_{LJ}=1.0\,\varepsilon/k_B$, where $k_B$ denotes the
Boltzmann constant.  The temperature was regulated via the
Nos\'{e}-Hoover thermostat~\cite{Allen87,Lammps}, and the periodic
boundary conditions were applied along all three spatial dimensions.
The computer glass transition temperature of the KA model at
$\rho=1.2\,\sigma^{-3}$ is
$T_c=0.435\,\varepsilon/k_B$~\cite{KobAnd95}. Following the
equilibration at $T_{LJ}=1.0\,\varepsilon/k_B$, the system was
cooled down to $T_{LJ}=0.01\,\varepsilon/k_B$ with the
computationally slow rate of $10^{-5}\varepsilon/k_{B}\tau$ at
constant volume. This preparation procedure was repeated for 100
independent samples.  Once at $T_{LJ}=0.01\,\varepsilon/k_B$, the
glass was strained along the $xz$ plane with the constant shear rate
$10^{-5}\,\tau^{-1}$ up to $\gamma_{xz}=0.20$. During strain
deformation, the potential energy, stress components, and atomic
configurations were periodically saved for further analysis.

\section{Results}
\label{sec:Results}


It is well realized by now that the mechanical response of
disordered solids under applied strain involves a series of rapid
rearrangements of clusters of particles that in turn induce
long-range deformation fields~\cite{BarratRev18}.  In general, the
atomic level description of such processes is complicated due to the
spatial overlap of the deformation fields and a wide distribution of
the local yield stresses. More recently, this problem was addressed
by constructing elastoplastic models where a disordered system is
represented via a collection of interacting blocks with certain
elastic properties and a distribution of yield
thresholds~\cite{BarratRev18}.  In the present study, we revisit the
problem of startup deformation of amorphous materials using
molecular dynamics simulations and report our observations in terms
of nonaffine displacements of atoms in the vicinity of the yielding
transition.

\vskip 0.05in

\begin{figure}[t]
\includegraphics[width=12.0cm,angle=0]{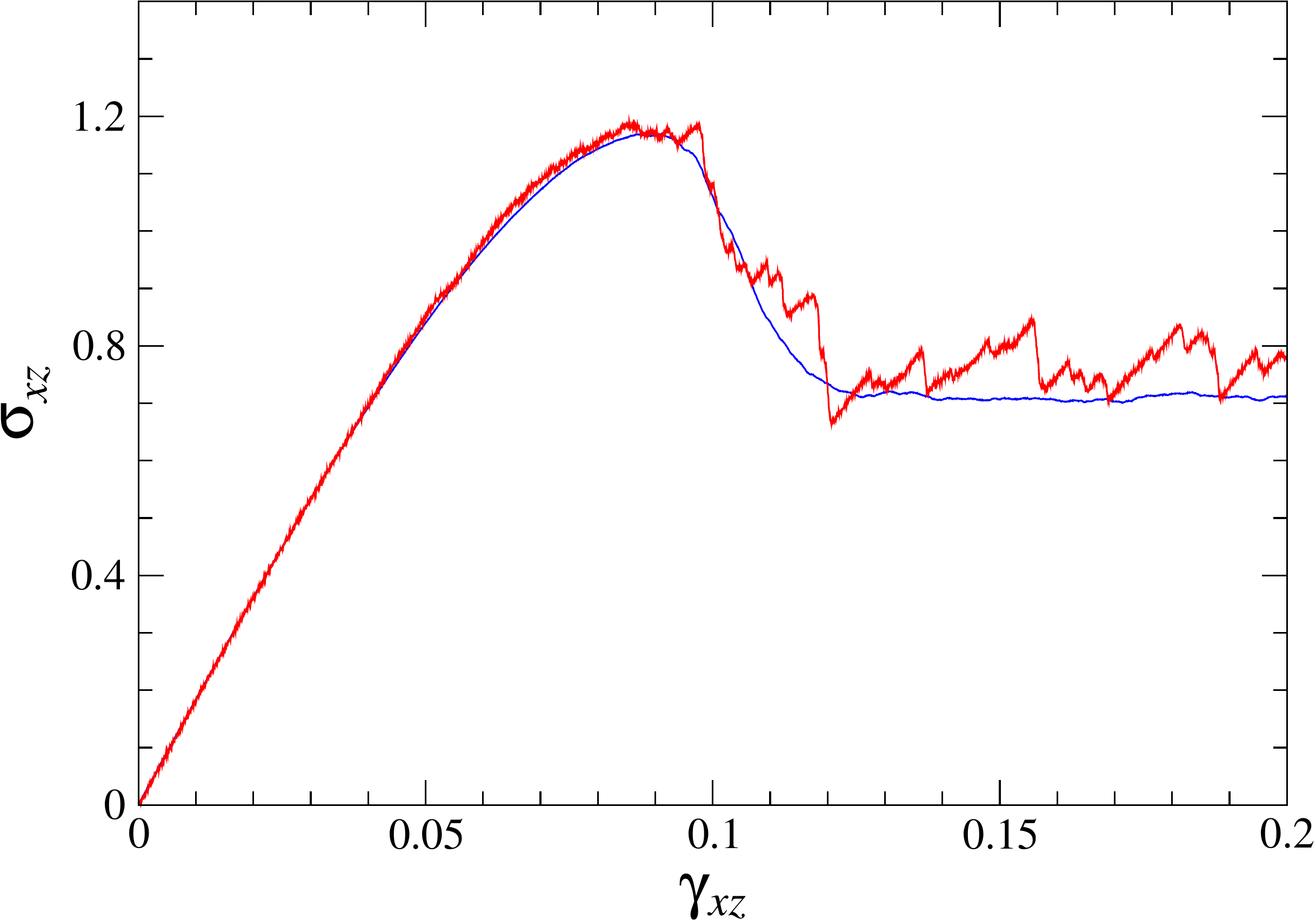}
\caption{(Color online) The shear stress $\sigma_{xz}$ (in units of
$\varepsilon\sigma^{-3}$) as a function of strain up to
$\gamma_{xz}=0.20$. The data are taken in one sample (the red curve)
and averaged over 100 samples (the blue curve). The shear rate is
$\dot{\gamma}_{xz}=10^{-5}\,\tau^{-1}$ and temperature is
$T_{LJ}=0.01\,\varepsilon/k_B$. }
\label{fig:stress_strain}
\end{figure}

The dependence of shear stress as a function of strain is presented
in Fig.\,\ref{fig:stress_strain} for the deformed glass with the
constant strain rate of $10^{-5}\,\tau^{-1}$. The data are collected
either in one sample (the red curve) or averaged over 100
independent samples (the blue curve). As is evident, the
stress-strain curve for one sample is punctuated by a series of
abrupt stress drops that are characteristic of sudden rearrangements
of clusters of atoms. In both cases, the stress response exhibits a
linear regime of deformation at small strain and a pronounced
yielding peak at about $\gamma_{xz}\approx0.09$, followed by a
nearly constant stress level at higher strain.  The appearance of
the stress overshoot is the signature of a relatively slow cooling
from the liquid state and/or an extended annealing period below the
glass transition temperature, when the glass is settled in a
sufficiently deep energy well. This behavior was repeatedly observed
in the previous MD
simulations~\cite{Varnik04,Horbach16,HorbachJR16,Sastry17}.

\vskip 0.05in


It should be commented that the value $\gamma_{xz}\approx0.09$ is
larger than the critical strain amplitude for yielding during
oscillatory shear deformation of the KA binary glass at the same
density and temperature~\cite{Priezjev17}. It was previous shown
that the critical strain amplitude of the KA mixture is in the range
$0.07< \gamma_0 < 0.08$ at the density $\rho=1.2\,\sigma^{-3}$ in
athermal quasistatic (the limiting case of $T_{LJ}=0$ and
$\dot{\gamma}_{xz}=0$) simulations~\cite{Sastry13,Sastry17}.
Moreover, it was demonstrated that the yielding transition occurs
after about 20 shear cycles at the strain amplitude $\gamma_0=0.08$,
when $\rho=1.2\,\sigma^{-3}$ and $T_{LJ}=0.01\,\varepsilon/k_B$ for
glasses prepared with the slow cooling rate
$10^{-5}\varepsilon/k_{B}\tau$~\cite{Priezjev17}.  The periodic
loading resulted in a gradual increase of the potential energy due
to accumulation of irreversible rearrangements, followed by the
formation of the shear band during three consecutive shear
cycles~\cite{Priezjev17}.  For reference, the oscillation period in
the previous study on the KA glass was set to $5000\,\tau$, and,
therefore, the shear rate at zero strain was equal to
$10^{-4}\,\tau^{-1}$~\cite{Priezjev17}. More recently, simulations
of the poorly-annealed binary mixture at higher temperature
$T_{LJ}=0.1\,\varepsilon/k_B$ and $\rho=1.2\,\sigma^{-3}$ have shown
that the yielding transition occurs after about 150 shear cycles at
lower strain amplitude $\gamma_0=0.06$~\cite{Priezjev18a}. Overall,
these results confirm that the critical strain amplitude is reduced
upon increasing
temperature~\cite{Priezjev13,Sastry17,Priezjev17,Priezjev18a}.

\vskip 0.05in

\begin{figure}[t]
\includegraphics[width=12.0cm,angle=0]{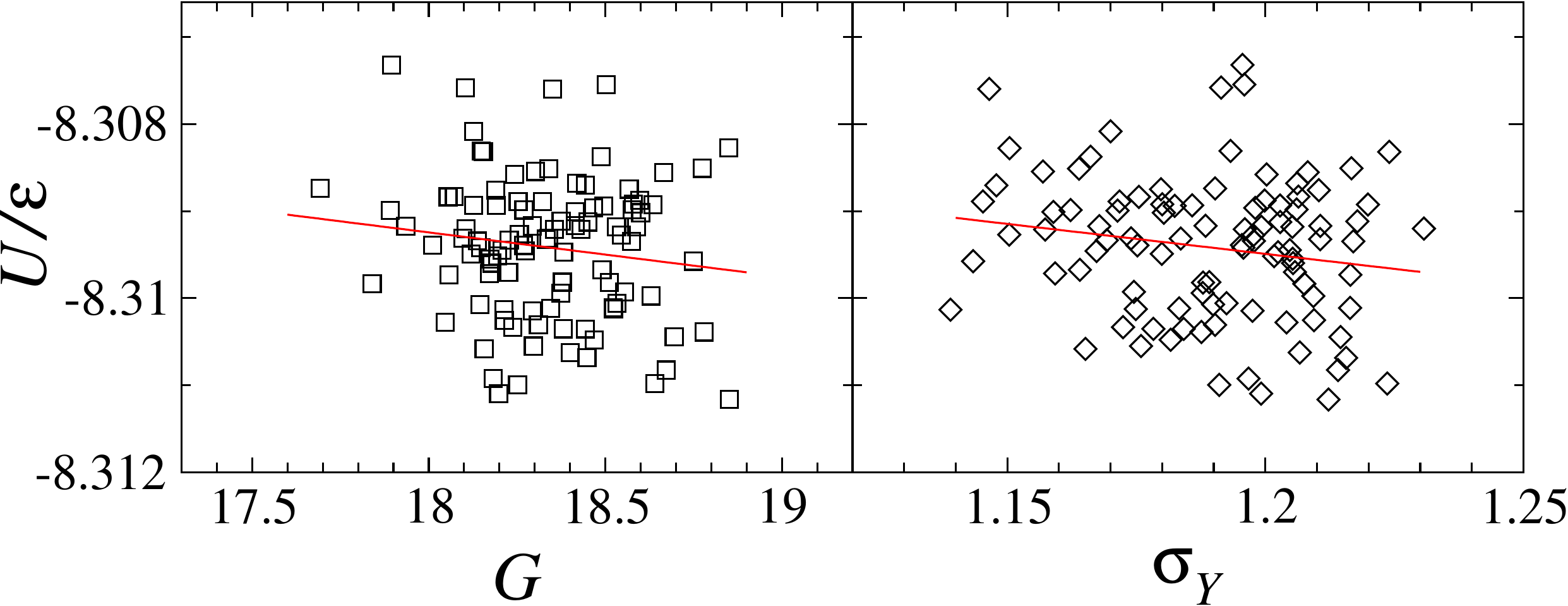}
\caption{(Color online) The shear modulus $G$ (in units of
$\varepsilon\sigma^{-3}$, left panel) and the yielding peak
$\sigma_Y$ (in units of $\varepsilon\sigma^{-3}$, right panel)
versus the potential energy at zero strain for 100 independent
samples. The red lines are the best fit to the data. }
\label{fig:poten_G_Y}
\end{figure}

Figure\,\ref{fig:poten_G_Y} shows the shear modulus and yielding
peak as a function of the potential energy at zero strain for 100
independent samples.   Here, the values of the potential energy are
reported for the samples cooled with the rate
$10^{-5}\varepsilon/k_{B}\tau$  to the temperature
$T_{LJ}=0.01\,\varepsilon/k_B$ at $\rho=1.2\,\sigma^{-3}$. The
variation of the data is related to different energy minima in the
potential energy landscape probed by the annealing of independent
samples. In turn, the shear modulus was computed from the linear
slope of the stress-strain curves at $\gamma_{xz}\leqslant0.01$,
whereas the height of the yielding peak was evaluated by averaging
the local maximum of the shear stress for each sample. The averaged
values of the shear modulus and the stress overshoot are
$G=18.3\pm0.2\,\varepsilon\sigma^{-3}$ and
$\sigma_Y=1.19\pm0.02\,\varepsilon\sigma^{-3}$, respectively. The
data are somewhat scattered but the trend can be clearly identified
via the negative slopes of the red lines that represent the best
linear fit to the data points. In other words, samples annealed to
lower energy states tend to have larger values of $G$ and
$\sigma_Y$.

\vskip 0.05in


The microscopic details of the deformation process in disordered
solids can be unveiled via the analysis of the so-called nonaffine
displacements of atoms. Unlike crystalline materials, where the
displacement of individual atoms can be defined with respect to the
periodic lattice, the description of atomic rearrangements in
amorphous materials is based on the relative displacements of atoms
with respect to their neighbors.  More specifically, the nonaffine
displacement of an atom $i$ is defined via the matrix
$\mathbf{J}_i$, which transforms the  neighboring atoms during the
time interval $\Delta t$ and, at the same time, minimizes the
quantity:

\begin{equation}
D^2(t, \Delta t)=\frac{1}{N_i}\sum_{j=1}^{N_i}\Big\{
\mathbf{r}_{j}(t+\Delta t)-\mathbf{r}_{i}(t+\Delta t)-\mathbf{J}_i
\big[ \mathbf{r}_{j}(t) - \mathbf{r}_{i}(t)    \big] \Big\}^2,
\label{Eq:D2min}
\end{equation}
where the summation is performed over atoms within the distance of
$1.5\,\sigma$ from the position of the $i$-th atom at
$\mathbf{r}_{i}(t)$. This quantity was first used by Falk and Langer
to unambiguously identify the local shear transformations in
deformed disordered solids~\cite{Falk98}.   Since then, the
spatiotemporal analysis of nonaffine displacements was extensively
applied to investigate the structural relaxation in amorphous solids
strained with a constant rate~\cite{Ma12,Chikkadi12,
Varnik14,Pastewka19,Priez19tfic} and periodically
deformed~\cite{Priezjev16,Priezjev16a,Priezjev17,Priezjev18,
Priezjev18a,NVP18strload,PriezELAST19,Priez19tcyc,PriezSHALT20}. In
general, it was found that subyield cyclic deformation leads to the
appearance of finite clusters of atoms with large nonaffine
displacements, and the size of these clusters is reduced with
increasing cycle number, when the system gradually approaches a
certain potential energy
level~\cite{Priezjev18,Priezjev18a,NVP18strload}. On the other hand,
at sufficiently large strain amplitude, the yielding transition
occurs via the formation of a shear band, typically after a number
of transient cycles~\cite{Priezjev17,Sastry17,Priezjev18a}.

\vskip 0.05in

\begin{figure}[t]
\includegraphics[width=12.cm,angle=0]{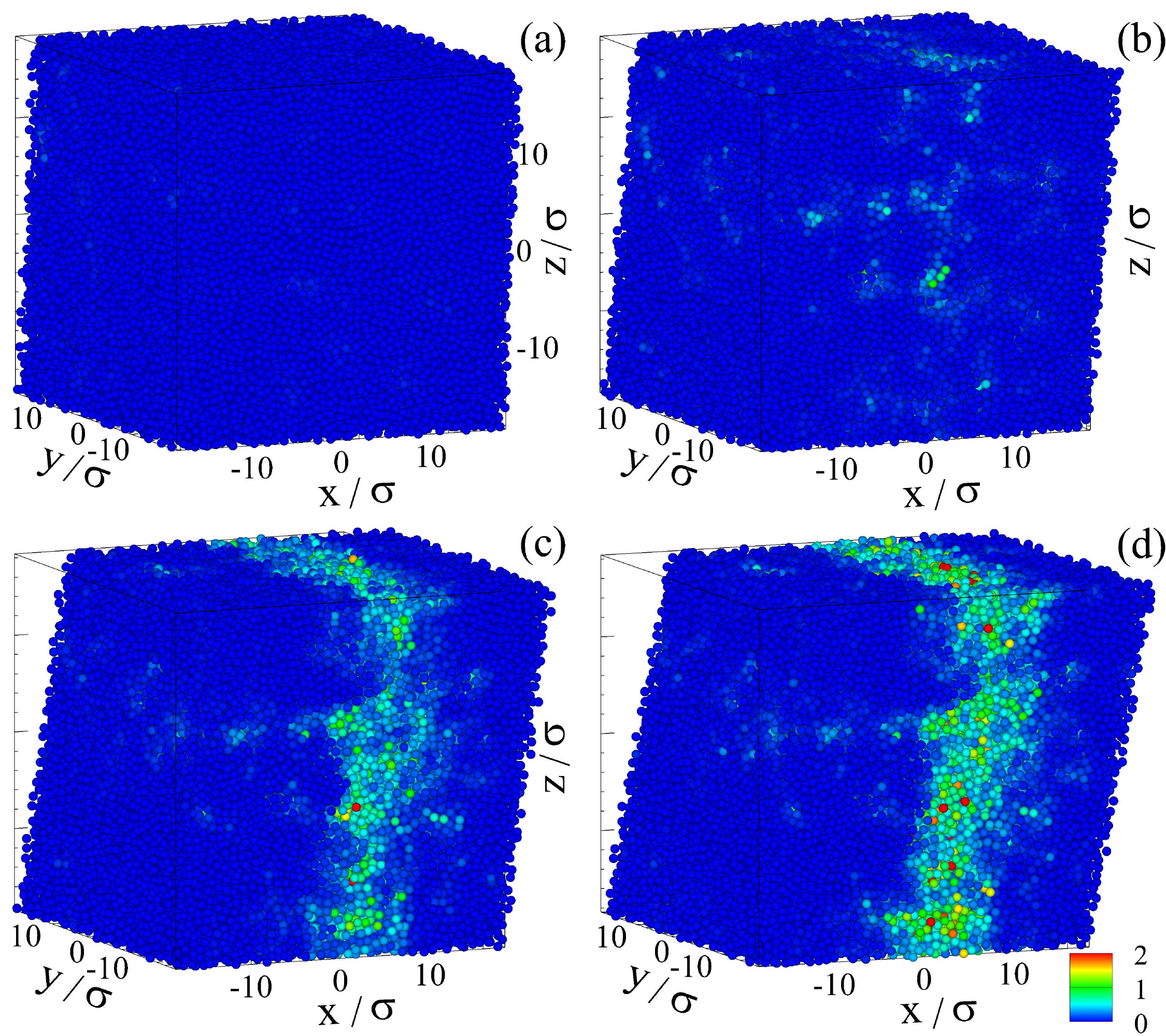}
\caption{(Color online) The sequence of snapshots of the binary
glass strained along the $xz$ plane with the rate
$10^{-5}\,\tau^{-1}$. The shear strain is (a) $0.05$, (b) $0.10$,
(c) $0.15$, and (d) $0.20$.  The color denotes $D^2$ with respect to
$\gamma_{xz}=0$, as indicated in the legend. Atoms are not shown to
scale. }
\label{fig:snap_gxz_20}
\end{figure}

The representative snapshots of the strained glass are displayed in
Fig.\,\ref{fig:snap_gxz_20} for the values of shear strain
$\gamma_{xz}=0.05$, $0.10$, $0.15$, and $0.20$. In each case, the
nonaffine measure, given by Eq.\,(\ref{Eq:D2min}), was computed with
respect to the atomic configuration at zero strain.  Note that the
snapshots in Fig.\,\ref{fig:snap_gxz_20} are taken for the same
sample as the stress-strain curve (the red curve) shown in
Fig.\,\ref{fig:stress_strain}. It can be clearly seen from
Fig.\,\ref{fig:snap_gxz_20} that shear deformation is initially
nearly homogeneous, and then it proceeds via the formation of the
shear band across the system with the width of several atomic
diameters. The unusual orientation of the shear band in this
particular sample is allowed due to the periodic boundary
conditions.

\vskip 0.05in

\begin{figure}[t]
\includegraphics[width=12.cm,angle=0]{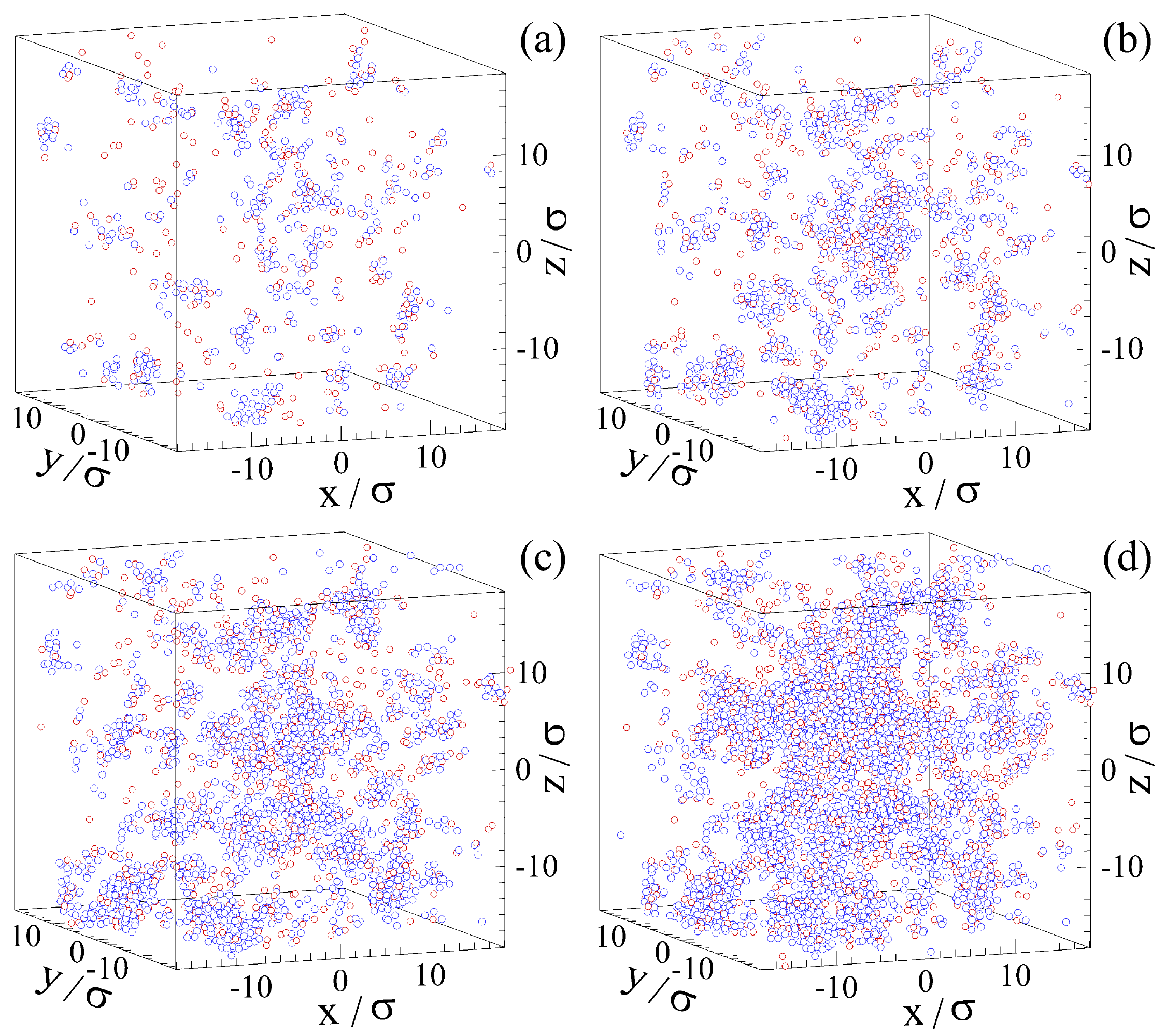}
\caption{(Color online)  The spatial configurations of atoms $A$
(blue spheres) and $B$ (red spheres) with the nonaffine measure: (a)
$D^2(0, 5\times10^3\tau)>0.04\,\sigma^2$, (b) $D^2(0,
6\times10^3\tau)>0.04\,\sigma^2$, (c) $D^2(0,
7\times10^3\tau)>0.04\,\sigma^2$, and (d) $D^2(0,
8\times10^3\tau)>0.04\,\sigma^2$.  The shear strain is (a) $0.05$,
(b) $0.06$, (c) $0.07$, and (d) $0.08$. The glass is strained along
the $xz$ plane with the rate $10^{-5}\,\tau^{-1}$.}
\label{fig:snap_D2_5_6_7_8}
\end{figure}
\begin{figure}[t]
\includegraphics[width=12.cm,angle=0]{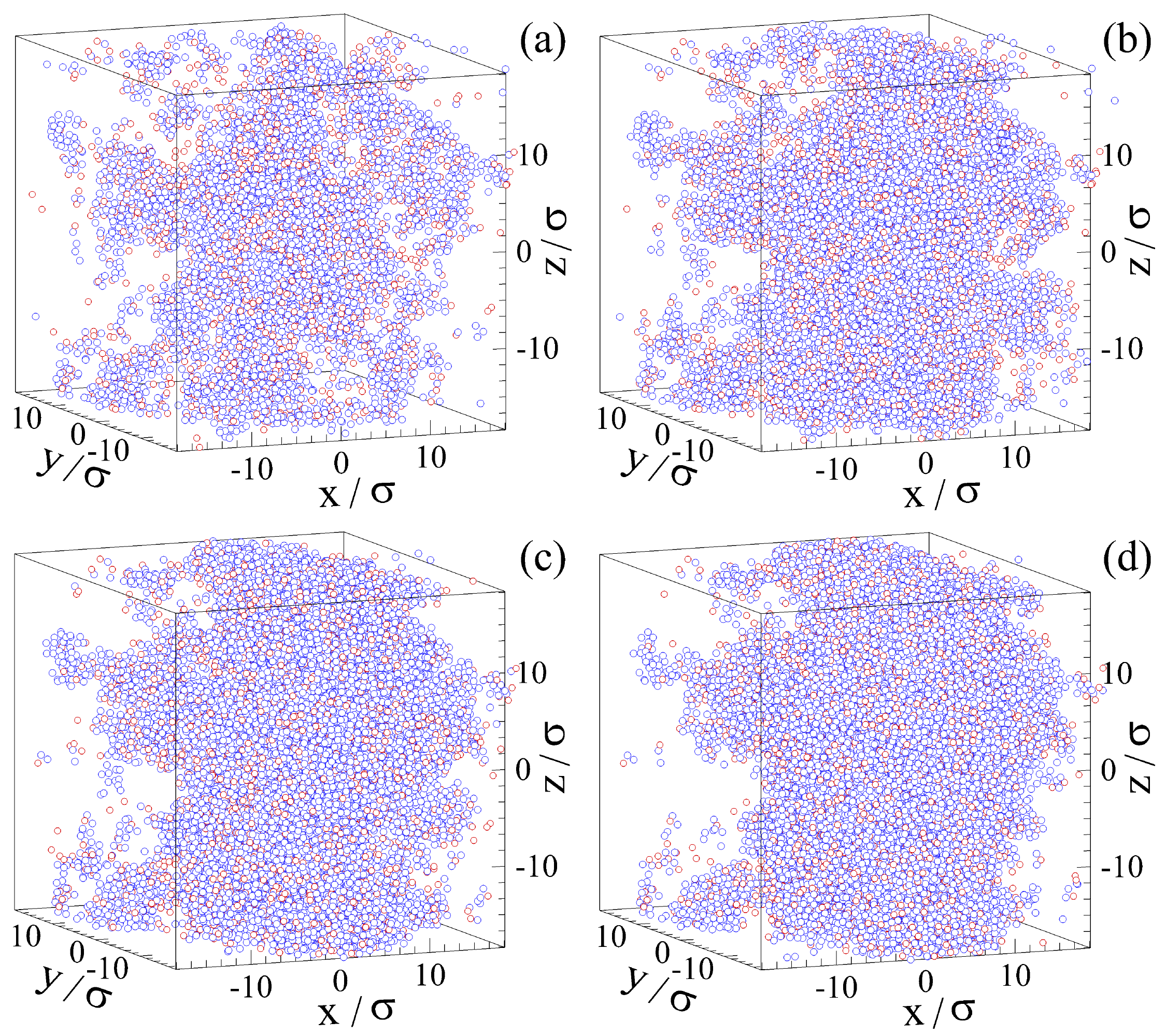}
\caption{(Color online)  The consecutive snapshots of sheared glass
at strains (a) $0.09$, (b) $0.10$, (c) $0.11$, and (d) $0.12$. The
nonaffine measure: (a) $D^2(0, 9\times10^3\tau)>0.04\,\sigma^2$, (b)
$D^2(0, 10\times10^3\tau)>0.04\,\sigma^2$, (c) $D^2(0,
11\times10^3\tau)>0.04\,\sigma^2$, and (d) $D^2(0,
12\times10^3\tau)>0.04\,\sigma^2$.   The shear rate is
$10^{-5}\,\tau^{-1}$.}
\label{fig:snap_D2_9_10_11_12}
\end{figure}

A more detailed analysis of spatial configurations of atoms with
large nonaffine displacements near the yielding point is shown in
Fig.\,\ref{fig:snap_D2_5_6_7_8} for strains $\gamma_{xz}=0.05$,
$0.06$, $0.07$, and $0.08$ and in Fig.\,\ref{fig:snap_D2_9_10_11_12}
for $\gamma_{xz}=0.09$, $0.10$, $0.11$, and $0.12$.  In
Figs.\,\ref{fig:snap_D2_5_6_7_8} and \ref{fig:snap_D2_9_10_11_12},
only the positions of atoms with $D^2(0, \Delta t)>0.04\,\sigma^2$
are displayed for clarity. The threshold value for the nonaffine
measure was chosen to be slightly larger than the typical cage size
$r_c\approx0.1\,\sigma$. It can be observed that finite clusters of
mobile atoms are already formed at relatively small values of
strain, \textit{i.e.}, $\gamma_{xz}\leqslant0.07$, see
Fig.\,\ref{fig:snap_D2_5_6_7_8}\,(a-c).  Furthermore, at the higher
strain $\gamma_{xz}=0.08$, the clusters of atoms form a larger
network, which is nearly homogeneously distributed in the sample,
see Fig.\,\ref{fig:snap_D2_5_6_7_8}\,(d).  Upon increasing strain,
the shear band forms along the plane at $x\approx 5\,\sigma$ when
$\gamma_{xz}=0.09$, as shown in
Fig.\,\ref{fig:snap_D2_9_10_11_12}\,(a). As evident in
Fig.\,\ref{fig:snap_D2_9_10_11_12}, the shear band becomes fully
developed at higher strain $\gamma_{xz}\geqslant0.10$.

\vskip 0.05in

\begin{figure}[t]
\includegraphics[width=12.0cm,angle=0]{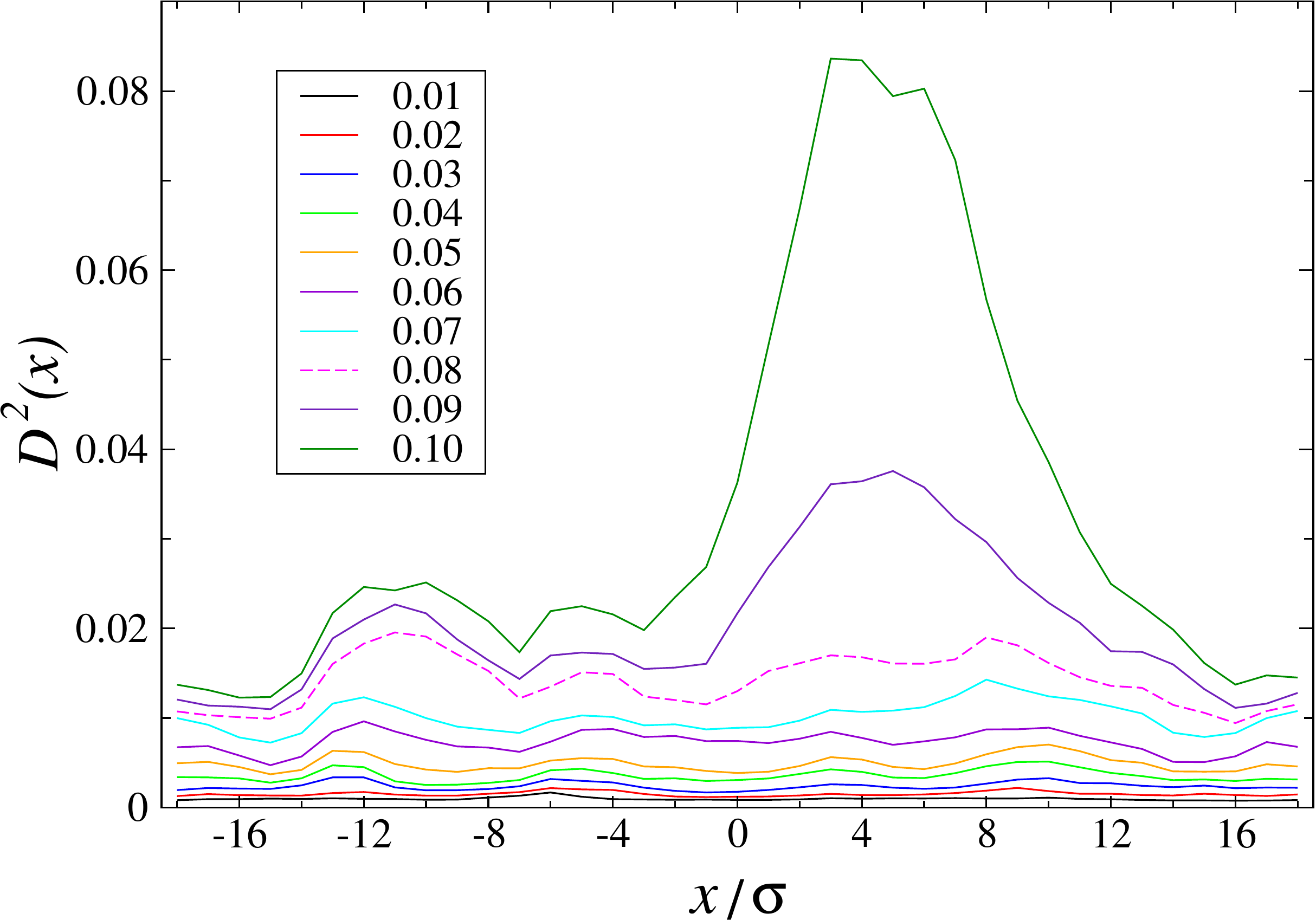}
\caption{(Color online) The spatial variation of the nonaffine
measure $D^2(x)$ for the indicated values of the shear strain. The
quantity $D^2(x)$ is computed in one sample with respect to an
atomic configuration at zero strain. See text for details. }
\label{fig:D2min_x}
\end{figure}

The quantitative analysis of the spatial and temporal evolution of
nonaffine displacements involves a sequence of spatially averaged
profiles of $D^2(x)$ within narrow bins of thickness $\Delta x =
\sigma$. The results are presented in Fig.\,\ref{fig:D2min_x} for
the values of shear strain $\gamma_{xz}\leqslant0.10$. It can be
clearly seen that the average level of $D^2(x)$ becomes higher when
$\gamma_{xz}$ increases, until $\gamma_{xz}=0.09$ when a pronounced
peak is developed reflecting the formation of the shear band. In
principle, one can assume that the location of a shear band can be
predicted from the spatial patterns of large nonaffine displacements
at shear strains smaller than the yielding point. The results in
Fig.\,\ref{fig:D2min_x} show, however, that generally this is not
the case. For example, the profile at $\gamma_{xz}=0.08$ (denoted by
the dashed curve in Fig.\,\ref{fig:D2min_x}) contains two local
maxima at $x\approx -10\,\sigma$ and $x\approx 8\,\sigma$. Despite
that the former maximum is slightly higher at $\gamma_{xz}=0.08$,
the shear band is formed at the location of the latter maximum when
$\gamma_{xz}=0.09$, see Fig.\,\ref{fig:D2min_x}. Altogether, these
results indicate that the yielding transition is associated with an
abrupt change from a nearly homogeneous distribution of clusters of
atoms with large nonaffine displacements to a localization within a
narrow region. Note that qualitatively similar conclusions were
obtained for both well and poorly annealed glasses subjected to
oscillatory shear deformation~\cite{Priezjev17,Priezjev18a}.

\vskip 0.05in

\begin{figure}[t]
\includegraphics[width=12.0cm,angle=0]{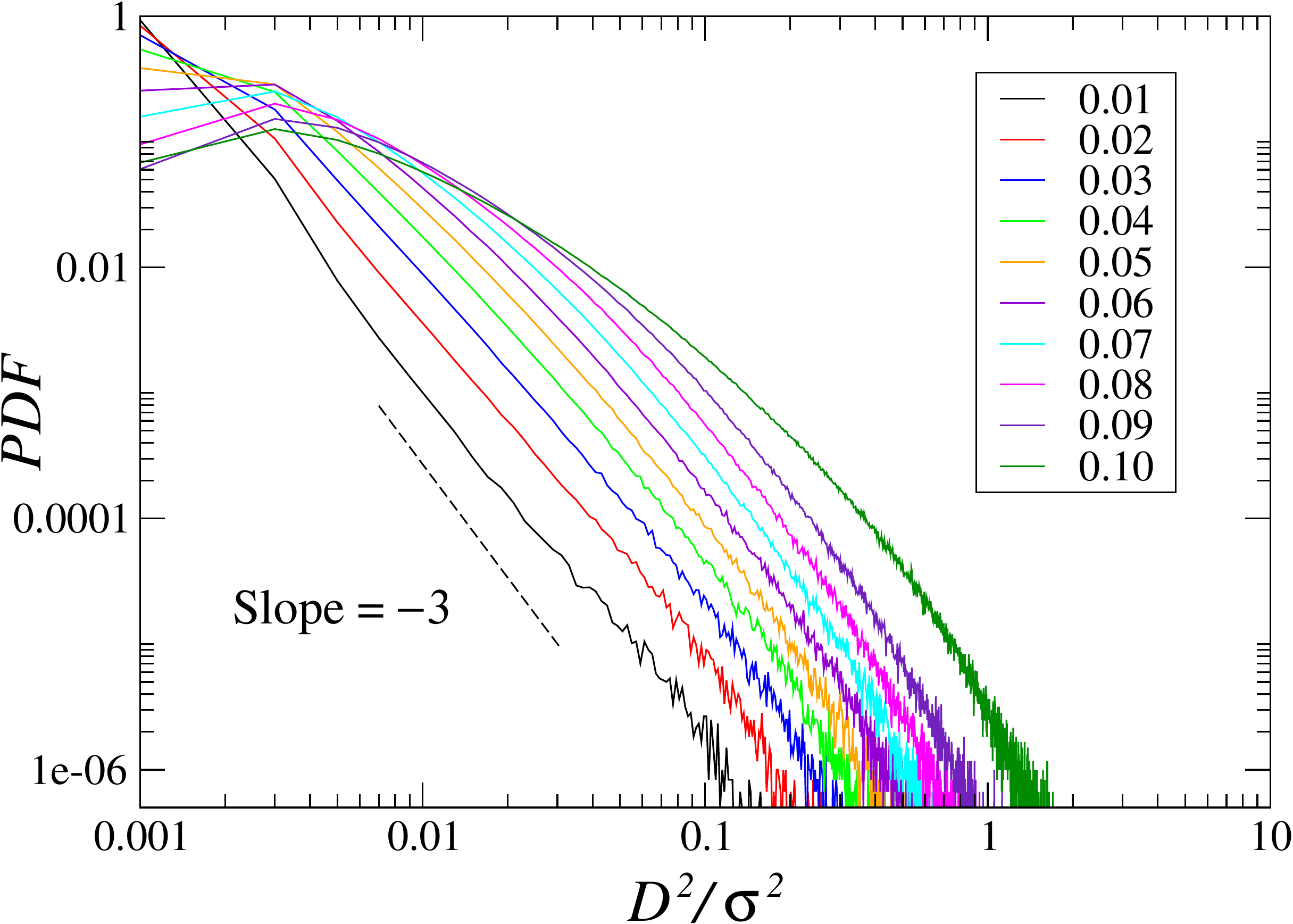}
\caption{(Color online) The normalized probability distribution
function (PDF) of $D^2$ for the listed values of shear strain. The
dashed line denotes the slope -3.  The data are averaged over 100
samples. }
\label{fig:PDF_D2}
\end{figure}

The normalized probability distribution function (PDF) of the
quantity $D^2$ is presented in Fig.\,\ref{fig:PDF_D2} for the values
of the shear strain in the range $0.01 \leqslant \gamma_{xz}
\leqslant 0.10$. As in the previous analysis, the nonaffine
displacements for each sample were computed with respect to atomic
positions at zero strain. It can be noticed that even at small
strain $\gamma_{xz}=0.01$, some atoms undergo large nonaffine
displacements, $D^2\approx 0.1\,\sigma^2$, and thus leave their
cages. As expected, the distribution function becomes wider with
increasing shear strain.  The relative increase in width becomes
larger at $\gamma_{xz}=0.10$, which is consistent with an abrupt
increase in the peak height reported in Fig.\,\ref{fig:D2min_x} at
$\gamma_{xz}=0.10$.

\vskip 0.05in

\begin{figure}[t]
\includegraphics[width=12.0cm,angle=0]{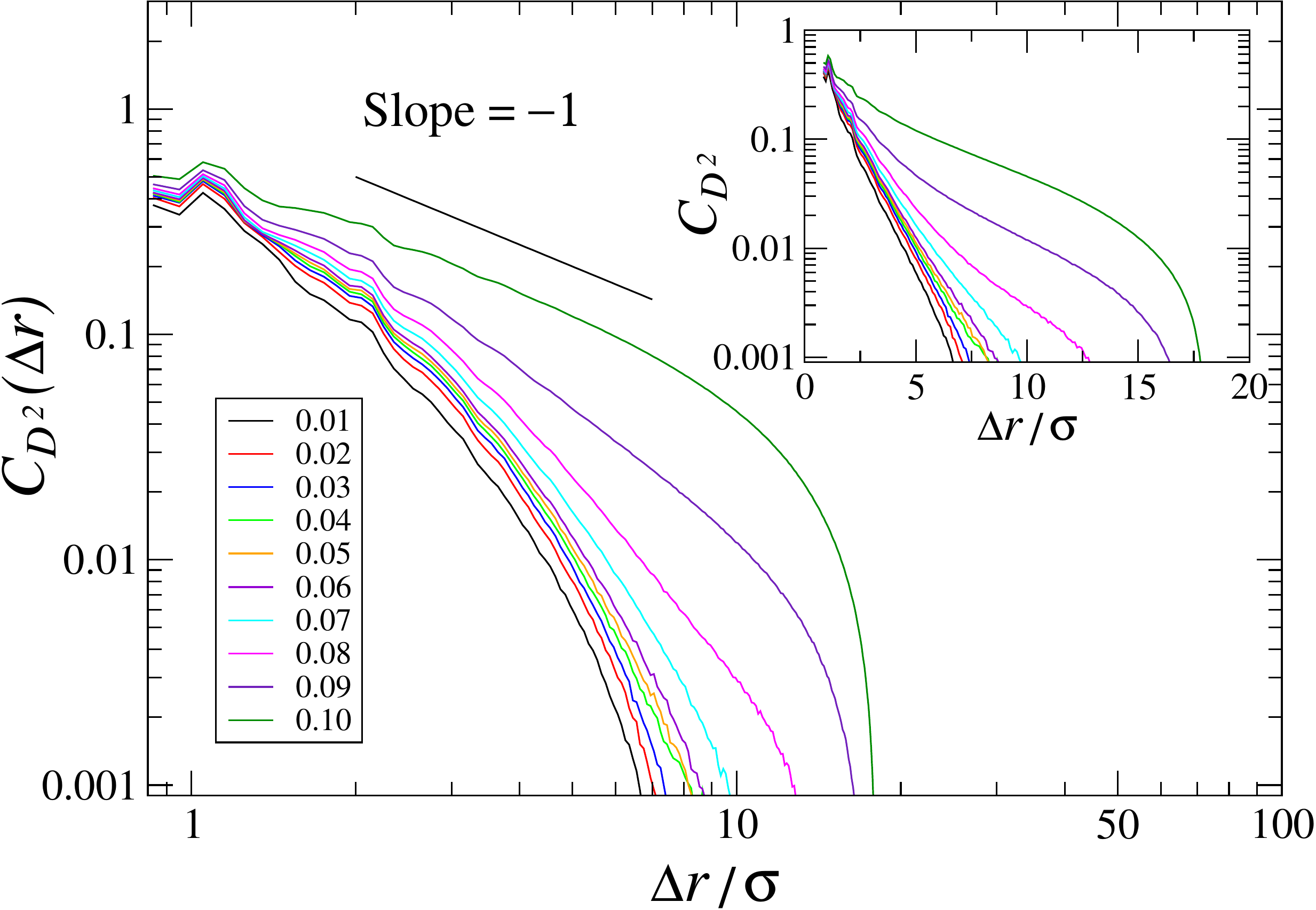}
\caption{(Color online) The spatial correlation function
$C_{D^2}(\Delta \textbf{r})$ defined by Eq.\,(\ref{Eq:CORR_D2}) for
the indicated values of shear strain. The straight line with the
slope -1 is shown for reference. Inset: The same data are replotted
on the log-normal scale. The data are averaged over 100 samples. }
\label{fig:C_D2}
\end{figure}

The spatial correlations of nonaffine displacements in a sheared
glass can be evaluated via the the normalized, equal-time
correlation function as follows:
\begin{equation}
C_{D^2}(\Delta \textbf{r}) = \frac{\langle D^2(\textbf{r} + \Delta
\textbf{r}) D^2(\textbf{r}) \rangle - \langle D^2(\textbf{r})
\rangle^2}{\langle D^2(\textbf{r})^2 \rangle - \langle
D^2(\textbf{r}) \rangle^2},
\label{Eq:CORR_D2}
\end{equation}
where the brackets indicate averaging over all pairs of
atoms~\cite{Chikkadi12,Varnik14}.  In our study, the correlation
function was further averaged over 100 independent samples for each
value of shear strain.  The results are shown in
Fig.\,\ref{fig:C_D2} for strains $\gamma_{xz}\leqslant0.10$. It can
be observed that the correlation of nonaffine displacements becomes
increasingly long ranged at higher strain. The decay of the
correlation function changes from exponential at
$\gamma_{xz}\leqslant0.07$ (see the same data on the log-normal
scale in the inset to Fig.\,\ref{fig:C_D2}) to the power-law decay
with the exponent approaching $-1$ at $\gamma_{xz}=0.10$.   The
tails of the power-law decay are strongly affected by the finite
system size, suggesting that simulations of larger systems might
show long range correlations more clearly. Nevertheless, it can be
concluded that upon increasing strain up to $\gamma_{xz}=0.08$, the
longer correlation range is consistent with larger size of typical
clusters of mobile atoms (\textit{e.g.}, see
Fig.\,\ref{fig:snap_D2_5_6_7_8}). Furthermore, the formation of the
shear band is associated with a relatively large increase in
$C_{D^2}(\Delta \textbf{r})$ when $\gamma_{xz}$ increases from
$0.08$ to $0.09$ (see Fig.\,\ref{fig:C_D2}).   We also comment that
it was previously shown that thermal fluctuations in quiescent
glasses lead to the exponential decay of $C_{D^2}(\Delta
\textbf{r})$, indicating that correlations of thermally-induced
nonaffine rearrangements extend up to nearest-neighbor
distances~\cite{Priezjev16a}.

\vskip 0.05in

\begin{figure}[t]
\includegraphics[width=12.0cm,angle=0]{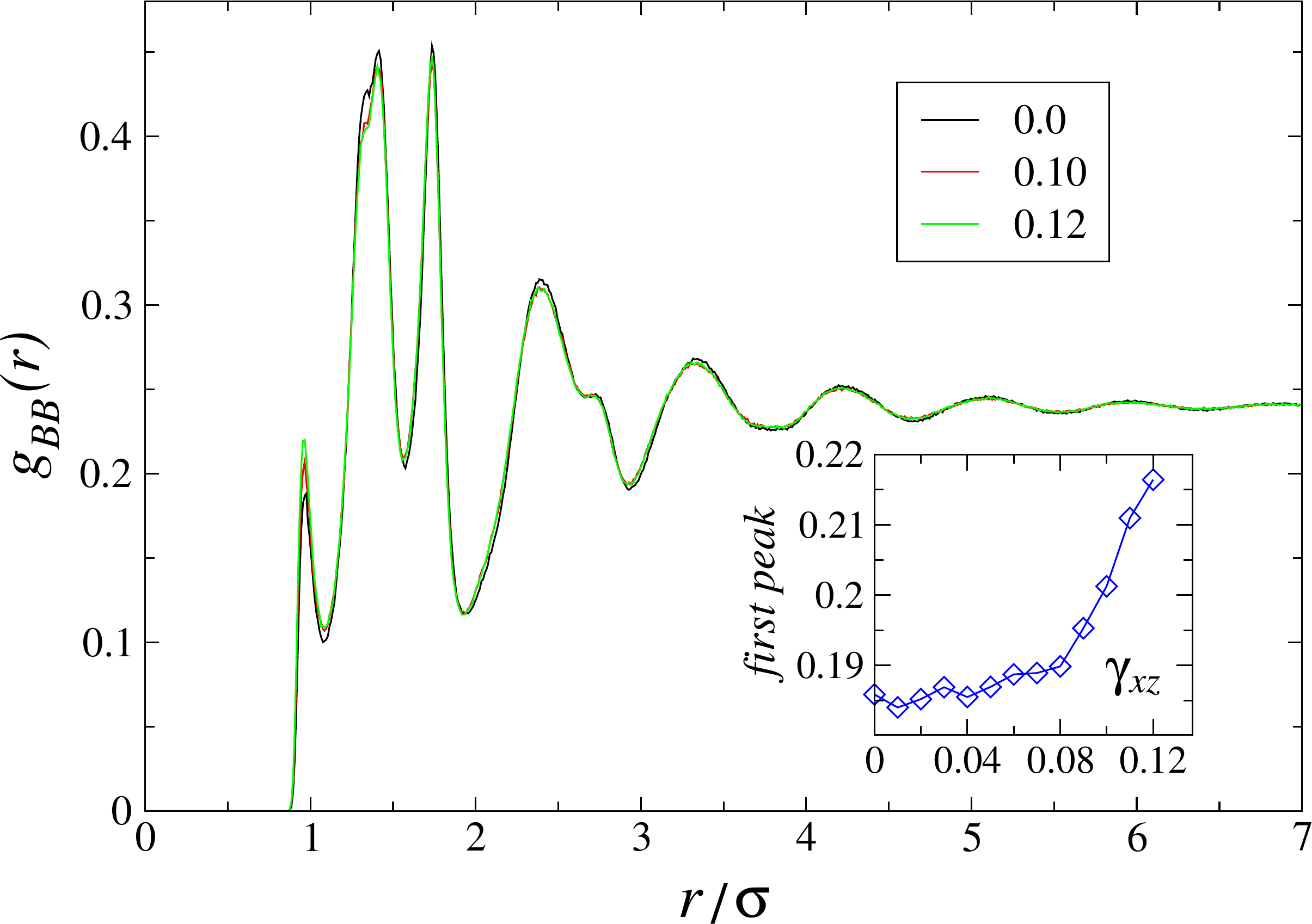}
\caption{(Color online) The pair correlation function, $g_{BB}(r)$,
for shear strain $\gamma_{xz}=0$, $0.10$, and $0.12$. The data are
averaged over 100 independent samples. The inset shows the height of
the first peak as a function of shear strain.}
\label{fig:gBB}
\end{figure}

We finally examine structural changes in binary glasses during
startup deformation. It was previously demonstrated that in contrast
to Zr-based metallic glasses, where the icosahedral short-range
order is an important structural measure, the atomic structure of
the KA binary glass is sensitive to the shape of the pair
correlation function of smaller
atoms~\cite{Stillinger00,Vollmayr96,Priez19one,Priez19tfic}. Recall
that the interaction energy $\varepsilon_{BB}$ in the LJ potential
is smaller than $\varepsilon_{AA}$ and $\varepsilon_{AB}$, and,
therefore, the probability of observing a neighboring pair of small
atoms of type $B$ is lower in well-annealed (low energy) samples.
The averaged pair correlation function, $g_{BB}(r)$, is presented in
Fig.\,\ref{fig:gBB} for strains $\gamma_{xz}=0$, $0.10$, and $0.12$.
As is evident, the height of the first peak is the lowest in
undeformed samples, and it becomes higher in strained glasses
($\gamma_{xz}=0.10$ and $0.12$) when atoms within a shear band made
several cage jumps.  Moreover, the dependence of the first peak
height as a function of strain is reported in the inset of
Fig.\,\ref{fig:gBB}. It can be seen that a distinct increase in
height coincides with the formation of the shear band at
$\gamma_{xz}=0.09$. This behavior is consistent with the development
of the main peak in the profile $D^2(x)$ at $\gamma_{xz}=0.09$ shown
in Fig.\,\ref{fig:D2min_x} and the strain localization reported in
Fig.\,\ref{fig:snap_D2_9_10_11_12}\,(a).

\section{Conclusions}

In summary, molecular dynamics simulations were carried out to study
the stress response and statistical properties of nonaffine
displacements during startup deformation of an amorphous solid. The
binary glass was prepared by slow annealing well below the glass
transition temperature at constant volume and then strained at
constant deformation rate. It was shown that atoms with nonaffine
displacements larger than the cage size form clusters that grow and
remain homogeneously distributed in space until the yielding
transition, which is associated with the formation of a localized
shear band.  This phenomenon is reflected in the variation of the
spatial correlations of nonaffine displacements, which change from
exponential to power-law decay near the yielding point. In addition,
the formation of a shear band coincides with a distinct increase in
the height of the first peak of the pair correlation function of
small atoms. Finally, the yielding transition during startup
deformation occurs at a value of shear strain greater than the
critical strain amplitude of oscillatory shear deformation at the
same density and temperature.

\section*{Acknowledgments}

Financial support from the National Science Foundation (CNS-1531923)
and the ACS Petroleum Research Fund (60092-ND9) is gratefully
acknowledged. The molecular dynamics simulations were carried out
using the LAMMPS code developed at Sandia National
Laboratories~\cite{Lammps}. The numerical simulations were performed
at Wright State University's Computing Facility and the Ohio
Supercomputer Center.


%
%

%

%

%

%

%

%

%

%

\bibliographystyle{prsty}

\end{document}